\documentclass[pra,twocolumn,showpacs,amsmath,amssymb,floatfix]{revtex4}

\pdfoutput=1

\usepackage[pdftex]{graphicx}
\usepackage{graphicx}

\newcommand{\be}{\begin{equation}}
\newcommand{\ee}{\end{equation}}
\newcommand{\bea}{\begin{eqnarray}}
\newcommand{\eea}{\end{eqnarray}}
\newcommand{\bma}{\begin{displaymath}}
\newcommand{\ema}{\end{displaymath}}

\begin{document}

\title{Many-particle dynamics of bosons and fermions in quasi-one-dimensional
flat band lattices}

\author{M. Hyrk\"as$^1$, V. Apaja$^1$ and M. Manninen$^2$}

\affiliation{$^1$Nanoscience Center, Department of Physics, FIN-40014 University
of  Jyv\"askyl\"a, Finland}
\affiliation{$^2$Rector's Office, FIN-40014 University
of  Jyv\"askyl\"a, Finland}

\date{\today}
 
\begin{abstract}

The difference between boson and fermion dynamics in quasi-one-dimensional lattices
is studied by calculating the persistent current in small quantum rings
and by exact simulations of the time-evolution of the many-particle state
in two cases: Expansion of a localized cloud, and collisions in a Newtons
cradle. We consider three different lattices
which in the tight binding model exhibit flat bands. The physical realization
is considered to be an optical lattice with bosonic or fermionic atoms. 
The atoms are assumed to interact with a repulsive short range interaction.
The different statistics of bosons and fermions lead to different dynamics.
Spinless fermions are easily trapped in the flat-band states due 
to the Pauli exclusion principle, which prevents them from interacting, while bosons
are able to push each other out from the flat-band states. 

\end{abstract}


\maketitle

\section{Introduction}          

Crystal structures existing in nature exhibit fascinating band structures which
determine electronic, optical, magnetic and thermal properties of materials. 
The curvature of an energy band determines the effective mass of the electron 
which can be hundreds of times the normal mass, like in heavy fermion 
materials \cite{stewart1984},
or infinite, like in the kagome lattice \cite{syozi1951} and other flat-band 
lattices \cite{deng2003,miyahara2005},
or even zero, like in graphene \cite{castroneto2009}.

Quantum dot lattices for
electrons \cite{lee2001,koskinen2003,reimann2002} and optical lattices
for atoms \cite{bloch2005,leggett2006,pethick2008,lopezaguayo2010} have
provided a new experimental setup where the lattice structure can be
formed artificially and, consequently, also structures which do not
exist in nature can be experimentally studied and utilized.  Moreover,
in optical lattices atom dynamics can be studied without problems
caused by lattice defects or phonons \cite{bloch2005,bloch2008,bloch2008b} and the
atoms trapped in the lattice can be chosen to be fermions or bosons.

The manipulation of the parameters of the optical lattice and the
properties of the atoms allows variation of the magnitude and even the sign
of the hopping parameters between neighboring lattice site \cite{struck-etal-2011}. 

In this paper we will study quasi-one-dimensional (Q1D) lattices with
flat bands. Our motivation is the fast development in the research of
atoms trapped in optical lattices, which has shown that surprisingly 
complicated lattice structures can be manufactured, such as the 
kagome lattice \cite{jo-2012}. Recently we suggested how a flat band exhibiting
2D edge--centered square lattice could be made \cite{apaja2010}. 
Such systems can be quite
accurately described with a Hubbard Hamiltonian with contact
interaction between the atoms. In the limit of an infinitely strong
interaction only one atom can occupy each lattice site and
consequently bosonic atoms will also have an "exclusion principle" in
the simple (localized) tight binding basis of single particle states.
However, bosons and fermions are different due to the different
symmetry of the many-particle wave function.  In this case of spinless
particles the fermion wave function is a single Slater determinant,
while the boson wave function is much more complicated consisting of
many permanents. This causes interesting differences in the quantum
dynamics of these two systems.  For example, the time evolution of
out--of--equilibrium bosons in 1D can be
non-ergodic \cite{kinoshita-2006}. The situation here is very different
from that of a rotating two-dimensional harmonic trap where the vortex
formation mechanisms in boson and fermion systems are closely
related \cite{saarikoski2010}. Trapped bosonic gas can be brought all
the way to the infinite repulsion limit (Tonks--Girardeau gas), where
bosons behave locally like free fermions - an effect known as fermionization
- but their momentum distribution is not
fermionic \cite{paredes-2004}. Flat bands add a very intriguing flavor
to the dynamics, as atoms occupying such bands are essentially immobile, 
and allow us to contrast bosons and fermions. In a magnetic field such immobility means  
localization in a so-called Aharonov-Bohm cage \cite{vidal-2000,doucot-2002}. 

In Section 2 we introduce three different Q1D lattices with flat
bands.  The simplest is a triangle lattice which has a flat band that is
separated by an energy gap from a normal band. In the diamond lattice the flat band
either cuts through two normal bands or can be separated from them. In
the stub lattice the flat band is between two normal bands that are
separated from it.

In Section 3 we introduce the many-body problem and describe how
the dynamical simulations were made and the persistent currents
calculated. In section 4 we describe the
results for the persistent current calculations, and in Section 5 the results for the
dynamical simulations. The conclusions are given in Section 6.

\section{Quasi-one-dimensional flat-band lattices}

There exists a large number of different lattice structures which in 
the simple tight-binding model, with only one state per lattice site and only 
the nearest neighbor hopping, exhibit band structures with one or more flat 
bands \cite{deng2003,miyahara2005,apaja2010}. 
Often the reason for the flat band 
is a solution where the single particle wave function is zero at some connecting sites 
of the lattice, making it impossible for the particles to move through the
lattice. This kind of lattices can be one, two or three dimensional. 

\begin{figure}[h!]
\includegraphics[width=\columnwidth]{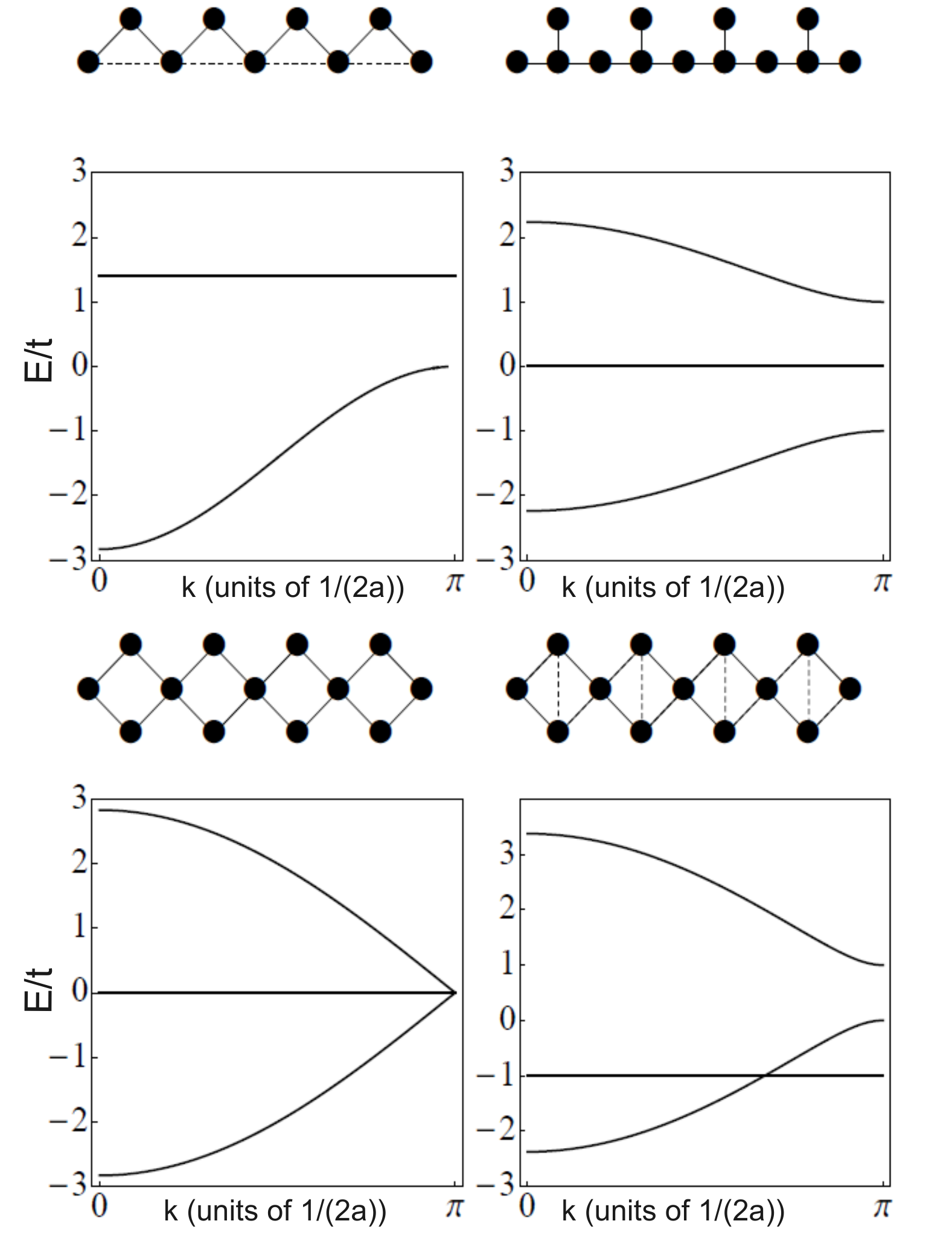}
\caption{Band structures of the Q1D lattices studied:
Triangle lattice, stub lattice, diamond lattice and diamond lattice with 
transverse hopping. The vertical axis shows the energy in units of $t=1$
and the horizontal axis the $k$-value in units of $1/(2a)$, $a$ being the lattice constant.
In the triangular lattice the hopping parameter shown as 
a dashed line has the value $t^\prime=1/\sqrt{2}$. The parameter for
the transverse hopping (dashed line) in the diamond lattice is $t^\prime=-1$.
}
\label{fig1}
\end{figure}

Fig.~\ref{fig1} shows the Q1D lattices and the corresponding band structures
studied in this paper. The single particle Hamiltonian is the H\"uckel-type
tight-binding (TB) model
\be
\hat{H}_{\rm TB}=-\sum_{i,j}^{\rm nn} t_{ij}\hat{a}_i^\dagger \hat{a}_j,
\ee 
where $\hat{a}_i^\dagger$ ($\hat{a}_i$) creates (destroys) a particle in lattice site $i$,
$t$ is the hopping parameter and 'nn' means that the sum is 
taken only over nearest neighbor sites. Throughout this paper
we use the natural units of the H\"uckel (or Hubbard) model where
$t=1$, $m=1$ and $\hbar=1$. 

In general, the hopping parameter $t_{ij}$ can vary in its
magnitude and sign from site to site. This can also be realized in
experiments, as was recently shown by Struck {\it et
  al.}~\cite{struck-etal-2011}. This independent tuning of the lattice
parameters allows one to move the flat band up or down in energy
in the diamond lattice, and to make the flat-band triangular
lattice discussed in this work.

In the triangle lattice the ratio
of the two hopping parameters has to be $t/t^\prime=\sqrt{2}$ in order 
to make one of the bands flat. For a positive $t$ the flat band is above
the normal band while for a negative $t$ it is below. 
In the stub lattice the hopping parameter is the same between all neighbors.
This lattice has a flat band in between two normal bands and
separated from them by gaps. The diamond lattice also has three bands
with the flat band in the center, but in this case all the bands meet at the 
Brillouin zone boundary. Adding a transverse hop between two points in each
diamond, as shown as a dashed line 
in the bottom right panel of Fig.~\ref{fig1}, moves the flat
band in relation to the other two bands. At the same time a gap opens between
the two normal bands. For positive values of the transverse hopping parameter
$t^\prime$ the flat band is raised, so that it either crosses the higher of the normal
bands or is above both of them. Similarly for negative values the band
is lowered (as show in the figure). Lowering the flat band can also be achieved
by flipping the signs of all the hopping parameters, which inverts the band structure.
This may be an easier configuration to realize in an experiment.

In the cases of the stub lattice and the diamond lattice the flat band appears
because the corresponding eigenstates have zero amplitude at the contact points of
the unit cells, preventing any motion of particles from one unit cell to another.
In the triangular lattice the explanation is more subtle, as will be seen
below when the persistent currents are considered.

\begin{figure}[h!]
\includegraphics[width=\columnwidth]{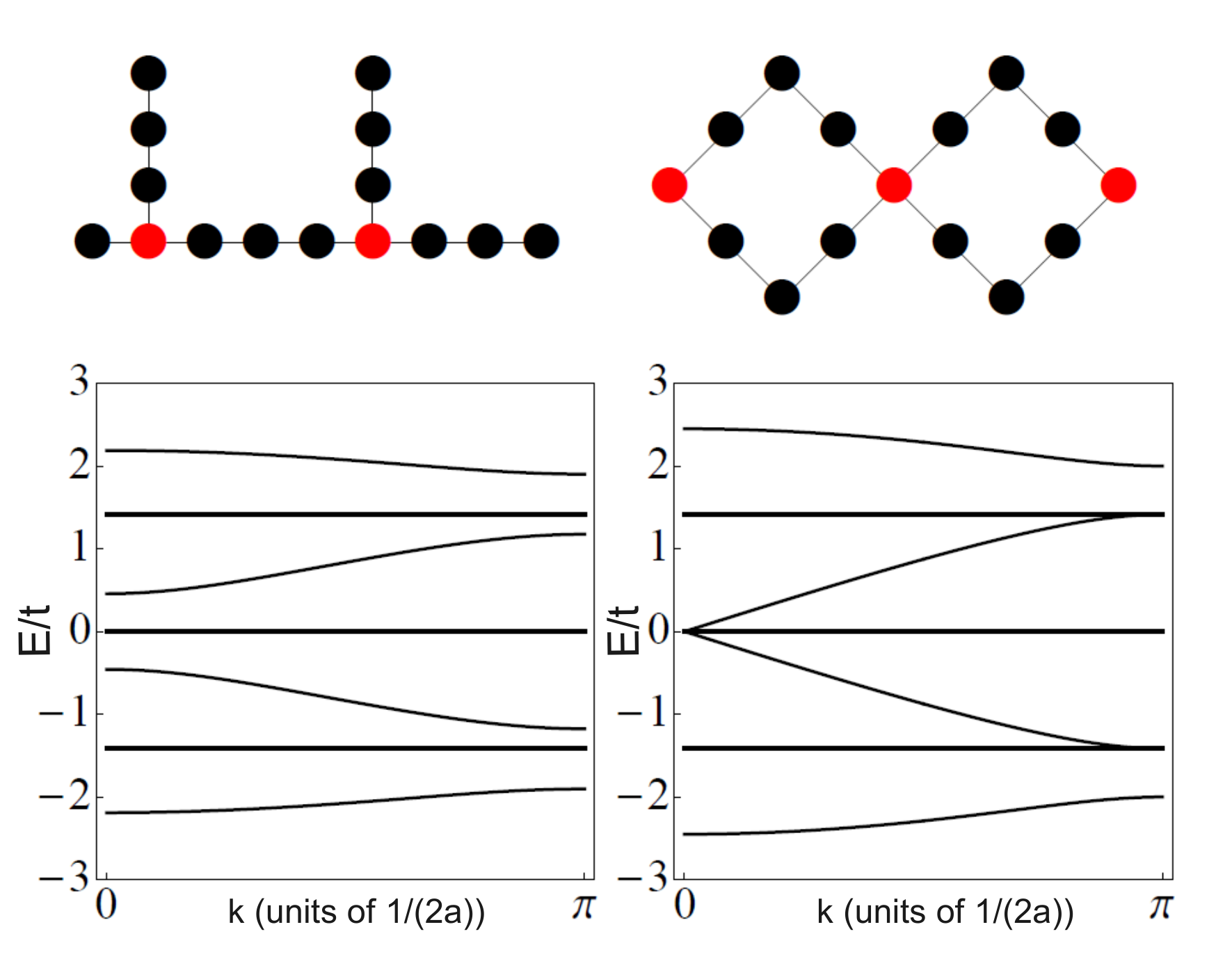} 
\caption{Band structures of extensions of the stub and the diamond
lattices. The vertical axis shows the energy in units of $t=1$
and the horizontal axis the $k$-value in units of the inverse lattice constant.
The gray nodal dots show the lattice sites at which the wave functions
of the flat-band states are zero.
}
\label{fig2}
\end{figure}

The stub lattice and the diamond lattice can be generalized to Q1D
lattices with several flat bands as demonstrated in Fig.~\ref{fig2}.
In each flat band the wave functions are zero at the corner points
shown as gray dots. The positions of the flat bands are then
determined by the length of the one-dimensional lattice (black dots)
between the corner points.  In both cases of Fig.~\ref{fig2} there are
three sites between the corner sites. These sites can be thought to
form a molecule with three sites and energy levels ($-\sqrt{2},\quad
0,\quad \sqrt{2}$) in the tight binding model.  These energy levels
determine the positions of the flat bands which are then naturally the
same in both lattices.

\begin{figure}[h!]
\includegraphics[width=\columnwidth]{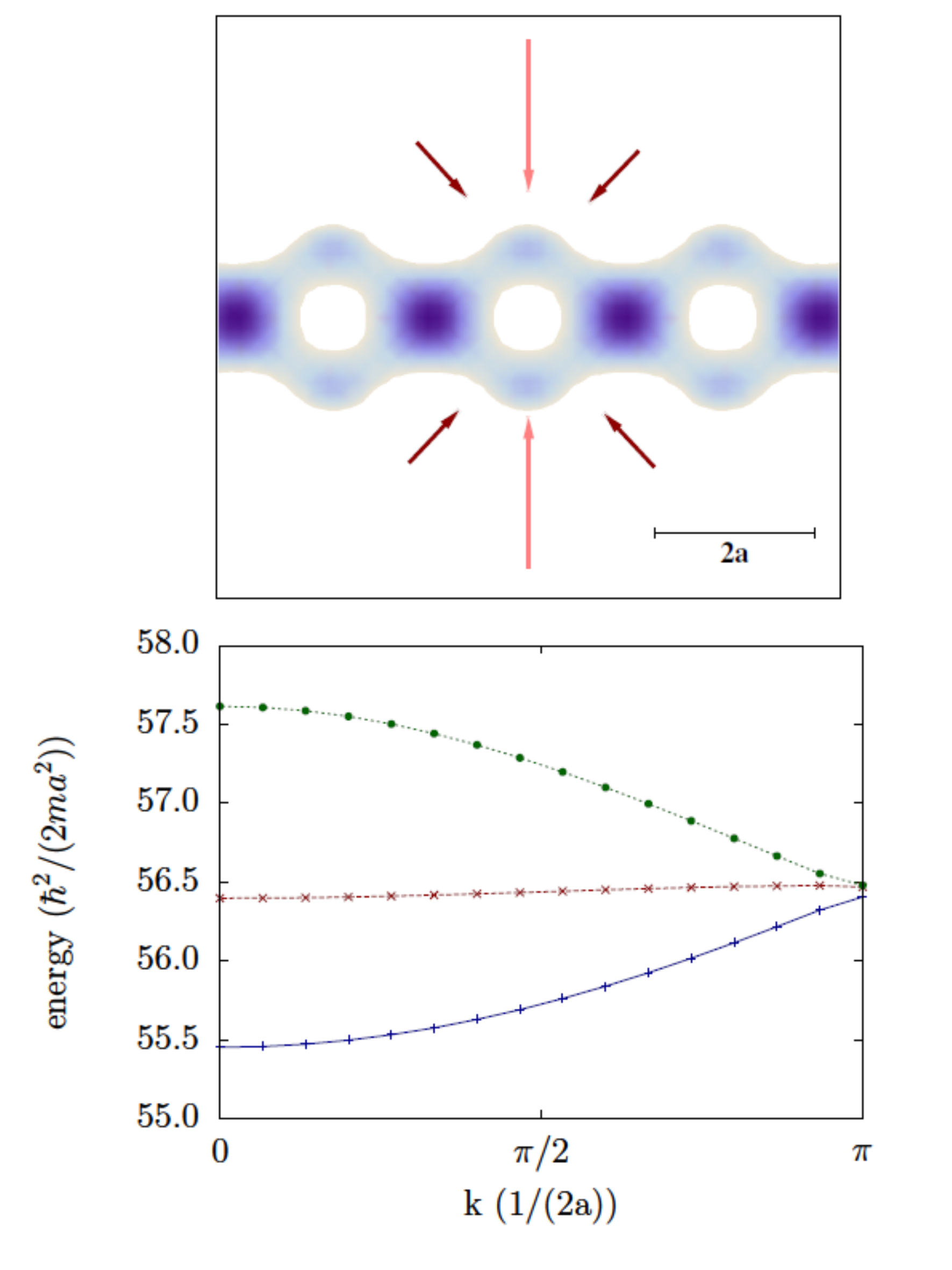} 
\caption{(Color online) Upper figure: Suggested laser setup (with a
  harmonic confinement in the vertical, $y$ direction) to produce a
  diamond chain lattice, and the resulting optical potential given in
  Eq.~(\ref{eq:Vdiamond}).  Darker areas denote lower potential.
  Lower figure: The three lowest tight-binding bands for the
  potential.  }
\label{fig:lasers}
\end{figure}

Next we consider how cold atoms in an optical lattice could be
arranged to a diamond lattice structure. In experiments, one induces a
dipole moment in the atoms with an oscillating electric light field
from a laser. Interaction of the dipoles with standing electromagnetic waves
creates the trapping potential for the optical
lattice, as reviewed in Ref.~\cite{bloch2005}. Multiple lasers can be arranged to trap
atoms in two- or three-dimensional optical lattices, where the
trapping potential of each laser can be described with a trigonometric
function, $\cos^2(x)$ or $\cos(2x)$. One possible laser setup to
create a linear diamond lattice is shown in Fig.~\ref{fig:lasers},
together with the optical potential. We solved the tight-binding band
structure for the potential
\begin{equation}
V(x,y) = V_0 [\cos^2(x+y)+\cos^2(x-y)+1.042\cos^2(y)+y^2]\ ,
\label{eq:Vdiamond}
\end{equation} 
where the last term is an extra harmonic confinement in the direction of 
$y$ axis and $V_0=34.3\; \hbar^2/(2ma^2)$, where $2a$ is the period
along the diamond chain. The harmonic confinement is added to reduce 
the periodic structure in the $y$ direction to quasi-one-dimensional.  
The Laplacian in the Schr\"odinger
equation was approximated with a five-point stencil.
These constants make the second band very flat, with
width of only 7 percent of the width of the lowest band
($0.073\;\hbar^2/(2ma^2)$ {\it vs.}
$0.95\;\hbar^2/(2ma^2)$). Fig.~\ref{fig:lasers} shows the three
lowest bands, obtained using 30 points in the $x$ direction and 60 in
the $y$ direction. With these choices, the band gap to the fourth band
is already $13\; \hbar^2/(2ma^2)$.

\section{Many-body dynamics}

We consider atoms in an optical lattice and assume that the
interaction between them has such a short range that it is only
effective when the atoms are located in the same lattice site.
We assume the atoms to be spinless or to be in the same spin state. 
In the case of fermionic
atoms this means that the system is spin-polarized and thus each atom
has the same $z$-component of spin.
The many-body Hamiltonian describing the system is the
Hubbard Hamiltonian
\be
\hat{H} =-\sum_{i,j}^{\rm nn} t_{ij}\hat{a}_i^\dagger \hat{a}_j+\sum_i V_i \hat n_i+
\frac{U}{2}\sum_i \hat n_i(\hat n_i-1),
\ee
where $\hat n_i=\hat{a}_i^\dagger \hat{a}_i$, $V_i$ is a local potential 
and $U$ is the strength of the 
contact interaction. 

In the case of the spinless fermions the interaction term is
irrelevant since the Pauli exclusion principle requires that 
each occupation $n_i$ is either zero or one. The many-body 
problem thus reduces to a single particle problem. 
The many-body state is a Slater determinant made out of the
single particle wave functions which are solutions of the 
tight binding model.

In the case of bosons the situation is more complicated.
If the repulsive interaction is infinitely strong we can
assume that the occupation of each site can not be more than one.
Unfortunately, this does not completely remove the complexity of the 
many-body problem like in the case of fermions. However,
we can still neglect the interaction term of the Hamiltonian 
by restricting the Fock space to those states which have 
an occupation of 0 or 1 in each lattice site. 

In solving the many-particle Hamiltonian in the case of bosons
we use the localized basis in a finite length of the lattice and
with a small number of particles. This restriction of the
computation to small systems has the advantage of  allowing
an exact diagonalization of the Hamiltonian. 
Alternatively, we can use the basis of matrix product states (MPS), where 
increasing the matrix dimensions increases the overlap with the exact state.
Large systems, where exact diagonalization is impractical, can still be 
approximated by MPS with reduced matrix dimensions.
Both the ground state representation as MPS, and 
the time evolution of the state can be described using the time-evolving
block decimation (TEBD) method by Vidal\cite{vidal2004}. In TEBD one 
chooses a target size of the matrices, and as time evolution increases
entanglement and thus expands the matrices, one truncates the matrices
back to target dimensions by throwing out the least important contributions.
What is important is determined by the means of repeated singular value or Schmidt decompositions.
At least for a short time the truncation error does not affect the results appreciably.
TEBD propagates time step by step, leaving also a 
controllable time-step error to the results. In small systems both of these methods
produced the same results. 

We are interested in the effect of the flat bands on the many-body
dynamics. To this end it is useful to know how the single
particle states are occupied in a given many-body state.
In the case of fermions this is trivial since the many-body state is
a single determinant of the single particle states.
In the case of bosons we can determine the occupations by changing
the basis from the localized basis ($\vert \alpha \rangle$), where the many-body solution
is $\vert\Psi\rangle=\sum_\alpha A_\alpha\vert \alpha\rangle$,
to the tight-binding (TB) basis ($\beta$), where   
$\vert\Psi\rangle=\sum_\beta B_\beta\vert \beta\rangle$.
($\vert \alpha \rangle$ and $\vert \beta \rangle$ are Slater determinants or 
permanents made of the single particle states).
The occupations can be found without resolving the coefficients $B_\beta$ 
by writing the creation operator of the TB basis as
$\hat{b}_j^\dagger=\sum C_{ji}\hat{a}_i^\dagger$,
where the coefficients $C_{ji}$ are obtained from the TB solution.
The occupation of a TB basis state $k$ is then
\be
n_k=\langle \Psi\vert \hat{b}_k^\dagger \hat{b}_k\vert\Psi\rangle
=\sum_{\alpha,\alpha^\prime}\sum_{i,j} A_\alpha^* A_{\alpha^\prime} C_{ki} C_{kj}^*
\langle \alpha\vert \hat{a}_i^\dagger \hat{a}_j\vert\alpha^\prime\rangle.
\label{statedist}
\ee

The time dependence of the many-body state
after a sudden change of the Hamiltonian (in our case the local potential)
can be determined by solving the ground state of the many-body problem 
(for the initial potential) and all the many-body states for the final potential 
and expanding the initial state as
\be
\vert\Psi\rangle=\sum_p D_p \vert \Psi_p\rangle,
\ee
where $\vert\Psi_p\rangle$ is the $p$'th time-independent energy eigenstate of
the final Hamiltonian. The time-dependence now follows from the time-dependencies
of the final states which are known:
\be
\vert\Psi(t)\rangle=\sum_p D_p e^{-i E_p t} \vert \Psi_p\rangle,
\label{eq:timedev}
\ee
where $E_p$ is the energy eigenvalue of the state $p$.
Note that in the case of fermions each final state is a Slater determinant.
It then follows that the time dependence of the many-body state can be 
determined by following the time dependencies of the individual single-particle states.

In principle, an initial local potential is actually not required, since one does not need to
know the initial Hamiltonian, but only the initial state. One may, for example, prepare
particles in certain lattice sites by any means conceivable, and then suddenly release them to 
follow the time evolution given by the final Hamiltonian.

\section{Persistent currents}

Persistent currents in quantum rings with a few fermions have been extensively 
studied, for a review see \cite{viefers2004}. 
In the case of bosons the early work was related to the research of
macroscopic systems of $^4$He \cite{reppy1964,grobman1996}, while lately
several studies of persistent currents in toroidal traps of bosonic atoms
have been reported \cite{lyandageller2000,isoshima2000,wang2007,ryu2007,bargi2010, lembessis2010}.

There are several ways to produce a toroidal trap for atom 
condensates \cite{lyandageller2000,hopkins2004, morizot2006,griffin2008,baker2009}.
In the case of finite quantum rings made of the Q1D lattices
considered, we induce an effective magnetic flux through the ring in order to
induce a current. Neutral atoms do not interact with the magnetic flux in the same way as electrons
in metallic or semiconducting quantum rings. However, laser fields can generate a
phase change which has the same effect as a magnetic flux \cite{mueller2004,amico2005,fetter2009}.
An effective flux can be created using rotationally symmetric Laguerre-Gaussian laser modes.

In the Hubbard model a flux piercing the ring will cause a phase shift
to the hopping parameter $t_{ij}$, changing it to $e^{i \Phi_{ij}} t_{ij}$.
In the case of the triangle lattice we have to notice that the phase shift
$\Phi_{ij}$ is twice as large for the hop along the long edge of the triangle
than along the short edges, i.e. the total phase shift is
independent of the path of the particle from one point to another.
In the case of the stub lattice the phase shift along the stub is zero.

The persistent current can be determined as the derivative of the 
total energy with respect to the flux or by computing the expectation value
of the current operator between two points:
\be
J=\frac{\partial E}{\partial \Phi} \quad \textrm{or}\quad 
J = \langle \hat{J} \rangle = \langle i\sum_{i,j}^{\rm nn}t_{ij} e^{i \Phi_{ij}}\hat{a}_i^\dagger \hat{a}_j \rangle.
\label{currentJ}
\ee

In the case of bosons with infinitely strong contact interaction
($U\rightarrow \infty$) and no on-site potentials ($V_i=0$)
the Hamiltonian is the same for particles and holes, i.e.
$\hat{H}=\sum \hat{a}_i^\dagger \hat{a}_j =\sum \hat{a}_j \hat{a}_i^\dagger$ since
the operators commute when $i\ne j$. This means that
the ground state energy and the 
persistent current are symmetric with respect to
particles and holes, irrespective of the symmetry of the 
single particle spectrum. The situation is different for 
fermions due to the anticommutation rule, which changes
the sign of the Hamiltonian for holes. Consequently,
in the case of fermions the many-body energy and 
the persistent current are symmetric with
respect to particles and holes only if the single 
particle spectrum is symmetric.

In the strictly one-dimensional case both boson and fermion systems
are exactly solvable via the Bethe
ansatz \cite{lieb1963,lieb1963b,yang1967,lieb1968}.  However, already
in the strictly 1D case bosons and fermions differ due to the
different symmetry of the wave
function \cite{manninen-reimann-viefers-2012}. In both cases the current
is a periodic function of the flux through the ring, but depending on
the number of particles the periodicity can have a different phase
for fermions and bosons. In the case of a zero flux, the lowest energy
state for any number of bosons has zero angular momentum while for
an odd number of spinless fermions the lowest energy state has a
finite angular momentum ($L=N/2$) \cite{viefers2004,lembessis2010},
resulting in a finite current with an infinitesimal flux.

In the strictly one-dimensional case spinless fermions can not pass each other.
The same is true for bosons interacting with an infinitely strong delta function 
interaction. The flat-band lattices are necessarily quasi-one-dimensional  
and thus more complicated. 

\begin{figure}[h!]
\includegraphics[width=\columnwidth]{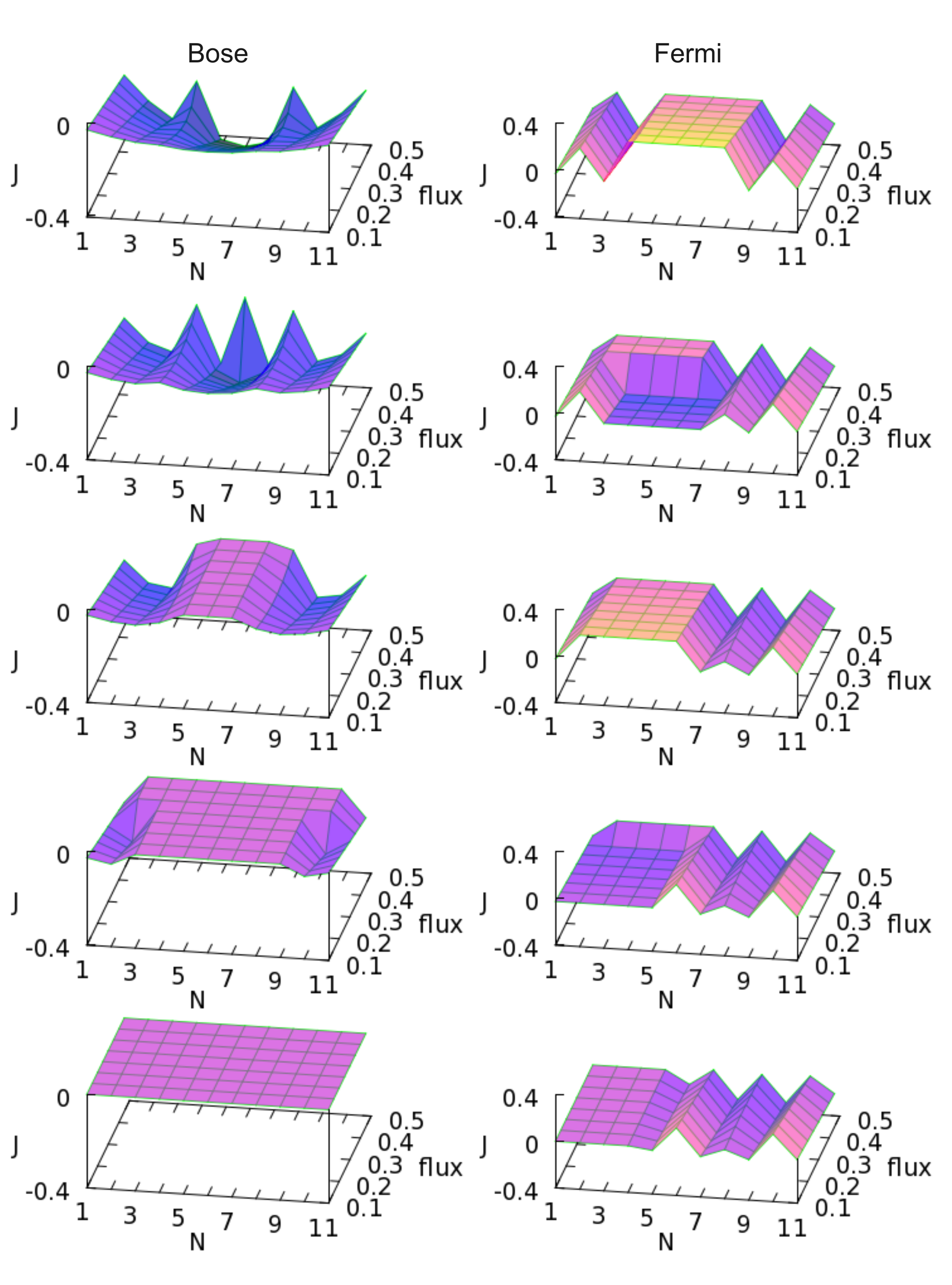}
\caption{(Color online) Persistent current of bosons (left) and fermions (right) in a
  quantum ring made of four unit cells of the diamond lattice.  The
  current $J$ is shown as a function of the number of particles in the
  ring ($N$) and the flux. In the upmost figures the flat band is in
  between the normal bands, and finally, in the lowest figures, below
  the normal bands (controlled by the value of the transverse hopping
  parameter, using values $t^\prime=0,-1.0,-1.5,-1.8,-2.1$). As the flat
  band moves down in energy bosons occupy it more and more, which
  finally leaves all the bosons immobile.  As particles are added the
  fermion currents sign alternates, except during the filling of the
  flat band when it remains constant. The stub lattice gives results
  similar to the upmost figures.}
\label{percur1}
\end{figure}

We study the persistent currents in the flat-band lattices by exactly diagonalizing
the Hubbard Hamiltonian for rings with a small number of lattice sites.
For bosons we assume an infinitely strong contact interaction
and for fermions we assume the system to be spin-polarized.
Fig. \ref{percur1} shows the results for rings made of the stub and the diamond lattices.
In each case the ring has 12 sites and from 1 to 11 atoms
(for 12 particles all the sites are occupied and no current can flow). 
We notice that the bosonic and the fermionic cases are markedly different.
The boson current shows the particle-hole
symmetry mentioned above, i.e. the result is the same for $N$ atoms and for $12-N$ atoms.

In the case of fermions the current is independent of the particle
number when the flat band is being filled. This is because the system
is noninteracting and the flat band can not conduct. The $N=2$ case
has a finite current already at zero flux. This is because in the
single particle picture this state has an angular momentum of 1 (or
-1) and thus a current. In the case of the stub lattice the current
for four particles and small flux is zero because the lowest band is
full.  In the case of the diamond lattice both normal bands meet the
flat band at the Brillouin zone boundary. Due to this degeneracy we
get a finite current for an infinitely small flux for particle numbers
$N=4\dots 8$.

\subsection{Diamond lattice}

The lowest panel in Fig.~\ref{percur1} shows the results for the diamond 
lattice where the transverse hopping shown in Fig.~\ref{fig1} has the value
$t^\prime=-2.1$, which brings the flat band below both of the normal bands.
In this case the persistent current for bosons is always zero. For fermions 
it is zero only for particle numbers $N=1\cdots 4$, which fit in the flat band.
For particle numbers 6 and 10 the ground state is degenerate and the current
starts from a finite value.  

In the case of bosons, the situation where the flat band is low in
energy is trivial, with most bosons occupying the flat band and
becoming immobile.  However, if the flat band is in the midst of the
normal bands the situation becomes more interesting. Looking at
Fig.~\ref{percur1}, one notices that the current is qualitatively
different for some atom numbers. The reason can be traced back to
rearrangement of single-particle energies and occupation.

\begin{figure}[h!]
\includegraphics[width=\columnwidth]{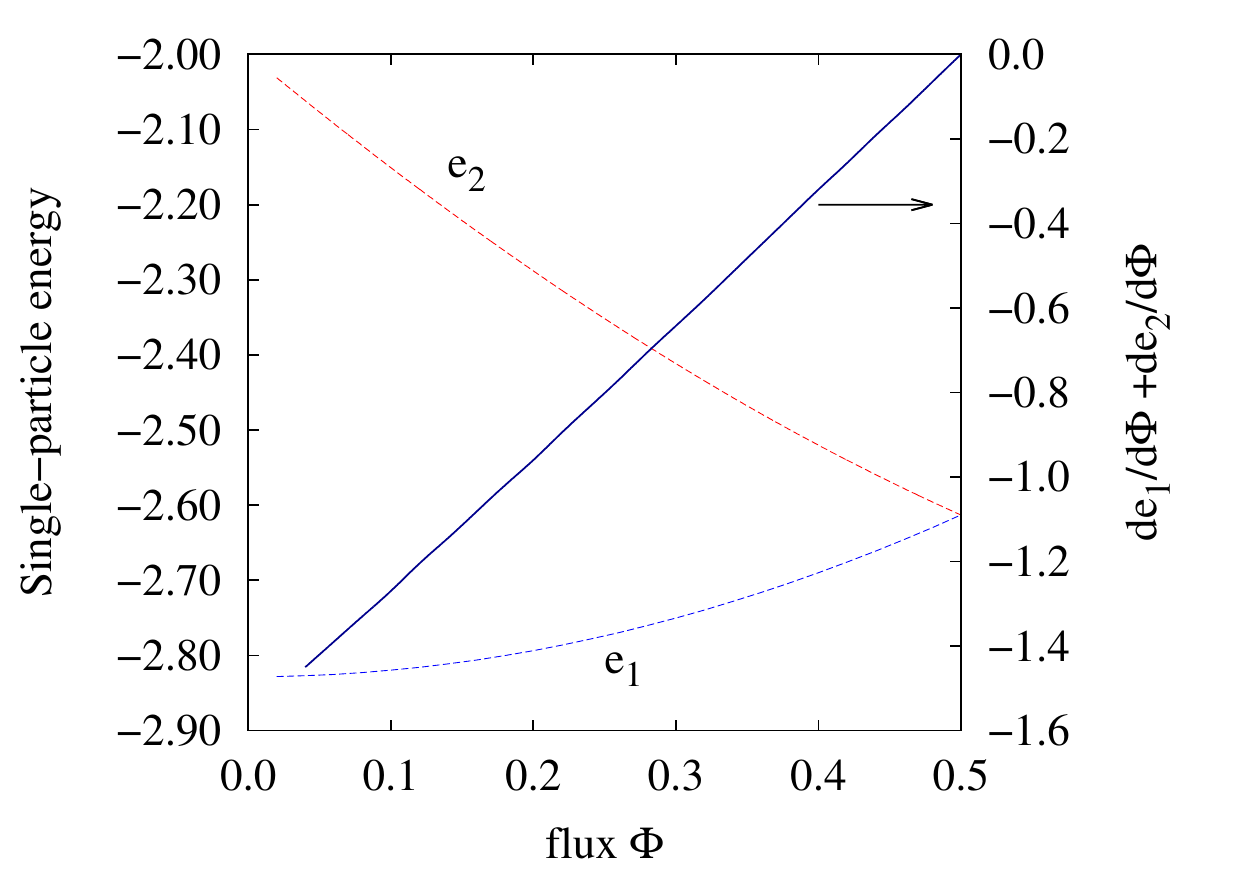}
\caption{(Color online) Energies of the two lowest single-particle states $e_1$ and
  $e_2$ (left $y$ axis) and the sum of the single-particle currents $de_1/d\Phi
  +de_2/d\Phi$ (right $y$ axis) as functions of flux.
  The figure shows, that currents carried by the single-particle states 1 and 2 
  are in opposite directions, and become equal in magnitude at flux $\Phi=0.5$.  
}
\label{singles}
\end{figure}

First of all, the infinitely repulsive on-site potential
does not show up while calculating the total energy of bosons with site
occupations of 0 or 1. Thus their total energy is simply the sum of the
single-particle energies, $E=\sum_i n_i e_i$, where $n_i$ are
obtained from (\ref{statedist}).  As a consequence, the
persistent current is the sum of ``single-particle currents'', using
(\ref{currentJ})
\begin{equation}
J=\frac{dE}{d\Phi}= \sum_i n_i \frac{de_i}{d\Phi} \ .
\end{equation}
Fig.~\ref{singles} shows that the single-particle currents
$\frac{de_i}{d\Phi}$ of states 1 and 2 circulate in opposite direction
(positive and negative slope, respectively), and become equal in
magnitude at flux $\Phi=0.5$, where the two states are also degenerate.
This is true also for state pairs (3,4),
(9,10) and (11,12). States 5-8 are on a flat band, and
don't carry any current.

\begin{figure}[h!]
\includegraphics[width=\columnwidth]{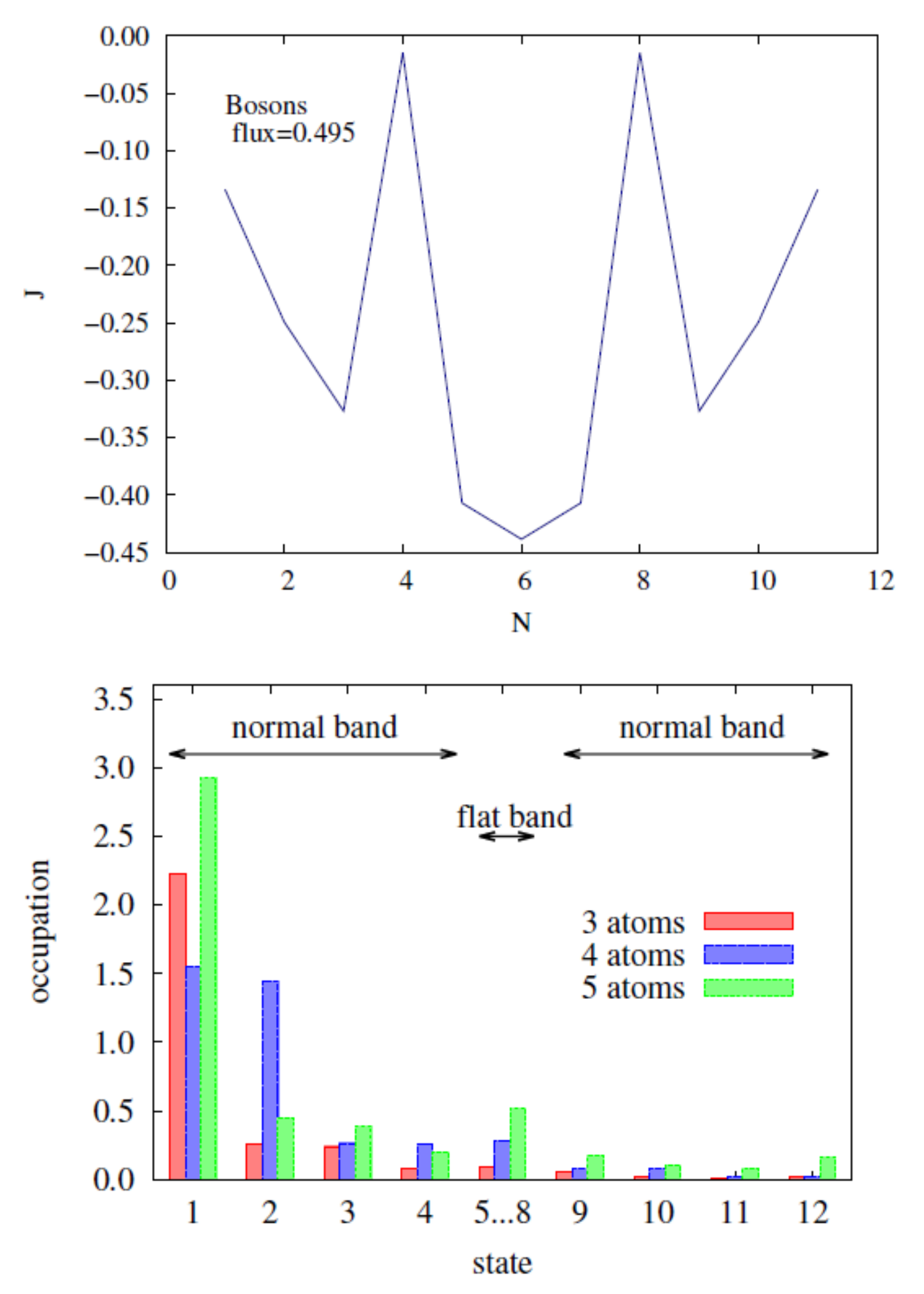}
\caption{(Color online) The upper figure shows the persistent current $J$ of bosons
  in a ring made of four unit cells of the diamond lattice as a
  function of the number of particles in the ring ($N$) for flux
  $\Phi=0.495$. At $N=4$ and $N=8$ the current is qualitatively different.
  The lower figure shows how the occupation of the single-particle
  states evolves as $N$ increases from 3 to 5. Flat-band states 5-8
  are degenerate. States 1 and 2 are almost degenerate, as are state
  pairs (3,4), (9,10) and (11,12).}
\label{percur1b}
\end{figure}

Next we want to see how the single-particle occupations change from
$N=3$ to $N=5$. We choose a flux slightly below 0.5 in order to see the
almost degenerate case. Upper panel of Fig. \ref{percur1b} shows in detail the
persistent current for flux $\Phi=0.495$. For this flux there is a very small
current at $N=4$ and $N=8$. Lower panel of Fig. \ref{percur1b} shows,
that the nearly vanishing current for 4 atoms corresponds to almost
equal occupations in the nearly degenerate state pairs (1,2), (3,4), (9,10)
and (11,12), so the opposite contributions to the current
cancel. Here the flat-band occupation is not directly
relevant. Another way to see why $N=4$ is special is to note that
there is exactly one atom per unit cell of three sites. The case $N=8$
has one hole per unit cell. Accordingly, for a diamond ring with
$R$ sites the boson persistent current vanishes at $N=R/3$ and
$N=R-R/3$.

\begin{figure}[h!]
\includegraphics[width=0.75\columnwidth]{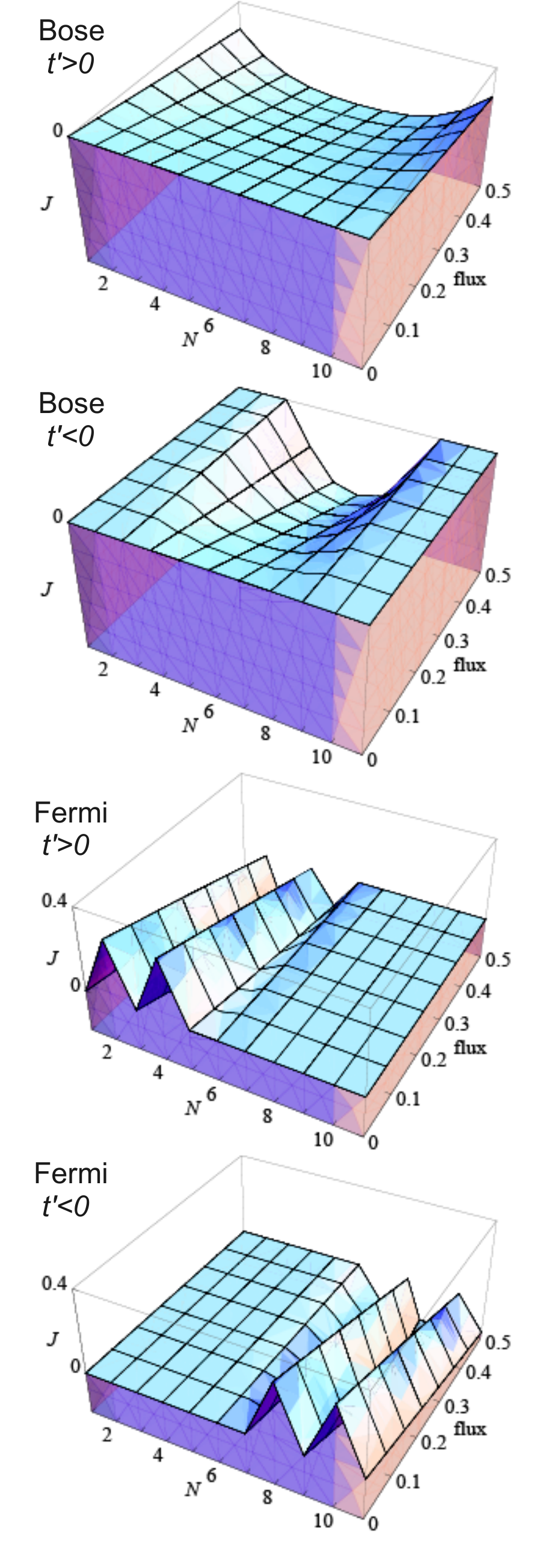}
\caption{(Color online) Persistent current of bosons (upper figures)
  and fermions (lower figures) in a quantum ring made of six unit
  cells of the triangle lattice.  The current (vertical coordinate) is
  shown as a function of the number of particles in the ring ($N$) and
  the flux. Note the different scale of the current for bosons and
  fermions.  In the case $t^\prime>0$  the flat band is above the normal
  bands, while in the case $t^\prime<0$ it is below
  the normal bands.}
\label{percur2}
\end{figure}

\subsection{Triangle lattice}

The results for the triangle lattice are shown in Fig.~\ref{percur2}.
We show results for cases where the flat band is
above the normal bands (positive $t^\prime$),
and where the flat band is below the normal bands (negative $t^\prime$). 
Note that the band structure does not depend on the sign of $t$. 

In both cases the current for bosons 
has a particle-hole symmetry, i.e. the current is the same for $N$ and 
$12-N$ particles, as discussed in the previous section. 
This symmetry causes a surprising effect
when the flat band is at the bottom:  The boson current
is zero for all flux values not only for small particle numbers ($N=1$, 2 and 3)
but also for large particle numbers ($N=9$, 10 and 11).

\begin{figure}[h!]
\includegraphics[width=\columnwidth]{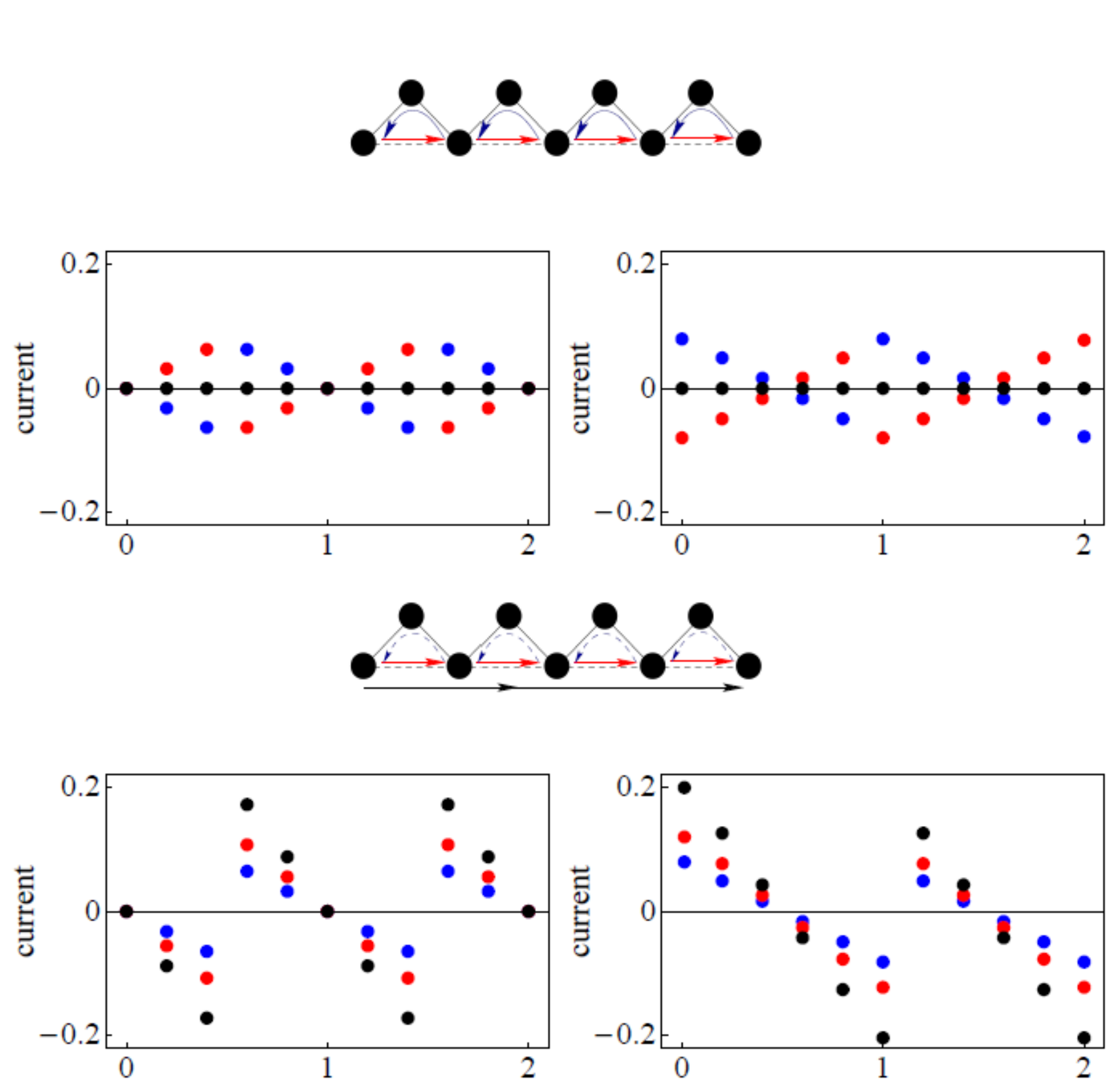} 
\caption{(Color online) Current components of two fermions and bosons in
a quantum ring of the triangular lattice. The black dots show the total current
and the red and blue dots the currents along the long edge and the short edge
of the triangle, respectively. The left panels show the results for bosons and the
right panels for fermions. The upper panels are for the case where the flat band
is below the normal band, resulting in zero net current. The lower panels are
for the case where the flat band is above the normal band.}
\label{tricur}
\end{figure}

The triangle lattice has the interesting feature that even when the total
persistent current is zero there is a current going around in each triangle.
This is demonstrated in the case of two particles in Fig. \ref{tricur}, 
which shows separately the currents going along the short 
edge and the long edge of the triangle. 
When the flat band is at the bottom the total current is zero, but 
the currents going along the short and long edges are nonzero with
opposite signs. This means that the flux, which is zero inside the triangles,
still induces a current going around each triangle. 

In the case where the flat band is at the top, the currents along the short 
and long edges of the triangles go to the same direction, but have different
magnitudes. Fig.~\ref{tricur} also demonstrates that the currents are 
periodic functions of the flux. In the fermionic case there is a discontinuity 
at integer flux values and in the bosonic case at half-integer flux values.
These discontinuities are caused by the degeneracies of the many-body 
states.

\section{Simulations of particle motion}

\subsection{Expansion of a localized cloud}

The results of the previous section show that for some particle
numbers the current through the ring is zero independently of the
value of the flux.  This suggests that the particles can be localized
at the flat-band states.  In order to study the localization further
we performed simulations of the dynamics. Initially the particles were
confined to a certain region of the lattice by adding a harmonic
confinement $V_{\rm h}(i)= \alpha i^2$, where $i$ is the distance from
the bottom of the harmonic confinement (in units of the distance
between the lattice sites along the ring) and $\alpha$ is the strength
of the potential. The unit of energy is $t$. Since we are interested
only in qualitative differences, we chose $\alpha=1$.

We studied the dynamics of four particles in a diamond lattice with 21 sites
and in a triangle lattice with 22 sites. In each case we solved the lowest energy
state of the Hubbard Hamiltonian with the harmonic confinement, and fully
diagonalized the Hamiltonian for the final state, i.e. without the harmonic potential.
This allowed us (using Eq.~(\ref{eq:timedev})) to study how the particles move when the harmonic 
confinement is suddenly removed. 

\begin{figure}[h!]
\includegraphics[width=0.75\columnwidth]{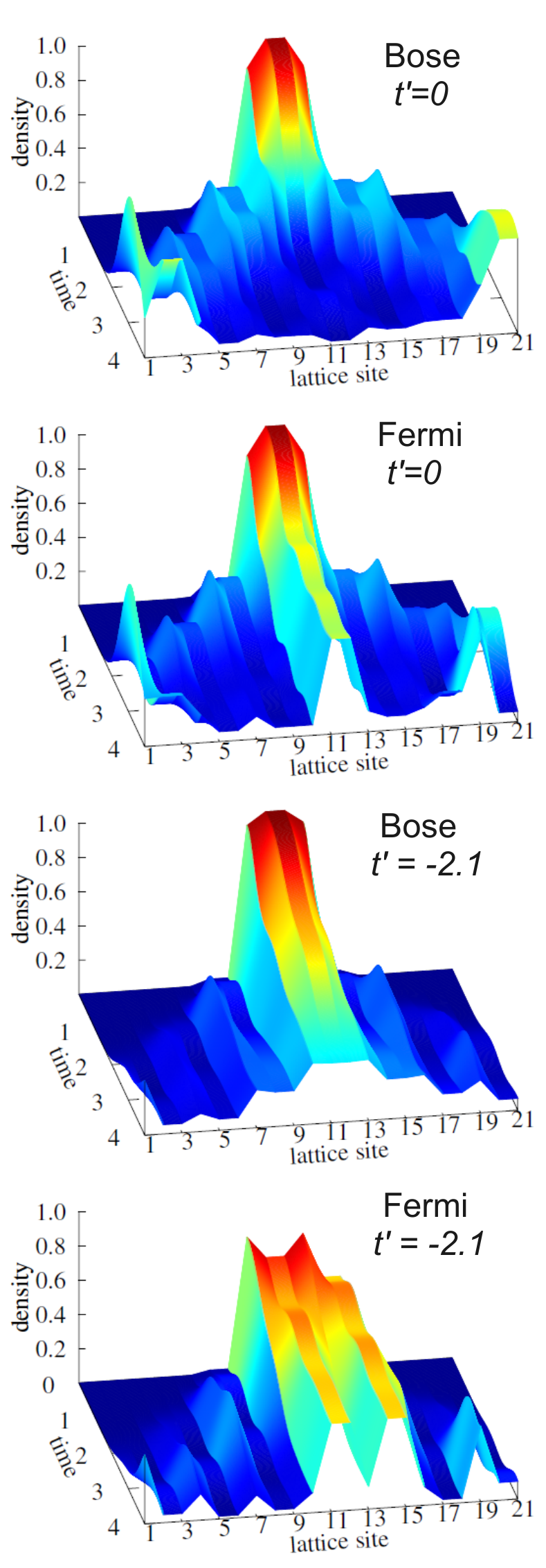} 
\caption{(Color online) Particle density (occupation) at different lattice sites of
  the diamond lattice as a function of time after the harmonic
  confinement is removed. The figures show the results for
  bosons and fermions in two cases: In the case $t^\prime=0$
  the flat band is in between the normal bands, and the
  in case $t^\prime=-2.1$ the flat band is below the normal
  bands.}
\label{evo1}
\end{figure}

Fig.~\ref{evo1} shows the results for the diamond lattice with 21
sites and periodic boundary conditions. The lattice sites were
numbered in succession so that the contact sites are 1, 4, 7, 10,
etc. and the two other sites in each diamond 2, 3, 5, 6, 8, 9, 11, 12,
etc. The center of the harmonic confinement was set at the sites 11
and 12, so that the potential was $V_{11}=V_{12}=0$,
$V_{10}=V_{13}=1$, $V_9=V_8=V_{14}=V_{15}=4$, and so on.  Initially
the particles are localized around the sites 11 and 12 using
confinement $V(x)=(x-x_0)^2$, where $x_0$ is the linear coordinate of
sites 11 and 12. It is convenient to use a linear coordinate along the
quasi-two-dimensional chain, so that sites at the tips of the
diamond are at the same potential. When the potential is removed
the particles start to move outwards until they reach the borders of
the simulation cell and start to overlap with the particles arriving
from the neighboring cells (due to the periodic boundary conditions).

The two upper panels show the results for the case $t^\prime=0$ where
the flat band is in between the normal bands. In this case all the
bosons become mobile and fly away, while some of the fermions stay at
the sites 11 and 12.  The situation is not much different when
$t^\prime=-2.1$ and the flat band is below the normal bands.  Also in
this case the bosons fly away, only more slowly.  The initial fermion
distribution is wider, but again some of the fermions stay
immobile. We also computed the dynamics for the stub lattice with 21
sites.  The results were qualitatively similar to those of the diamond
lattice with the flat band at the center.

\begin{figure}[h!]
\includegraphics[width=.8\columnwidth]{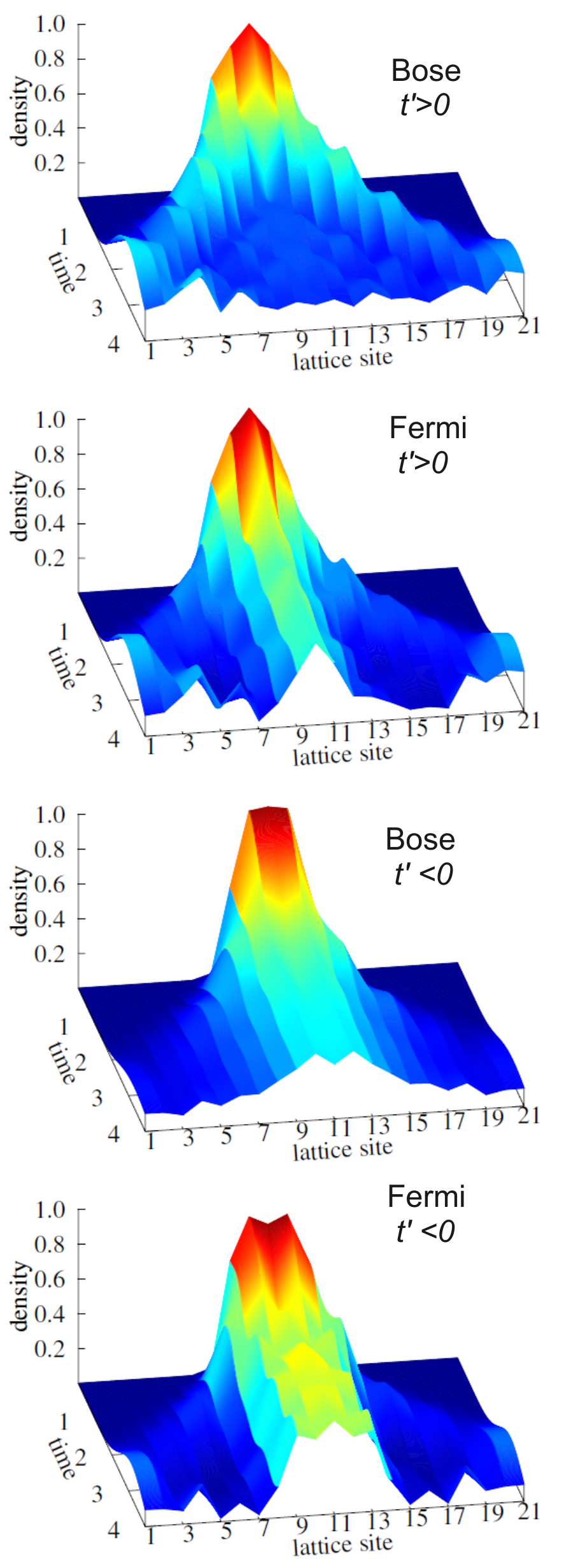}
\caption{(Color online) Particle density (occupation) at different lattice sites of
  the triangle lattice as a function of time after the harmonic
  confinement is removed.  The figures show the results for
  bosons and fermions in two cases: In the case $t^\prime>0$
  the flat band is above the normal bands, and the
  in case $t^\prime<0$ the flat band is below the normal
  bands.}
\label{evo2}
\end{figure}

Fig.~\ref{evo2} shows the results for the triangle lattice. In this
case we have 21 lattice sites and the center of the harmonic
confinement is at the site 11. The initial potentials are $V_{11}=0$,
$V_{10}=V_{12}=1$, $V_9+V_{13}=4$, $V_8=V_{14}=9$ and so on.  The two
upper panels of Fig.~\ref{evo2} show the case where the flat band is
at the top, and the two lower panels the cases where the flat band is
at the bottom. The results are rather similar to those for the diamond
lattice. In both cases the boson distribution widens with time and all
the bosons eventually fly away. Also in both cases a part of the
fermions stays localized.

We have repeated the dynamics simulations for different particle numbers and
numbers of lattice sites, using both periodic boundary conditions and 
a finite length lattice. In all cases the results are qualitatively
similar to those shown in Figs.~\ref{evo1} and \ref{evo2}.

The general result, that some of the fermions stay localized in the 
flat-band states, is easy to understand. Spinless fermions 
with contact interaction are equivalent to noninteracting fermions.
Those initially occupying a flat-band state will stay there when the 
confinement potential is removed. The same is not true for bosons,
which can push each other out from the flat-band states. 

\begin{figure}[h!]
\includegraphics[width=\columnwidth]{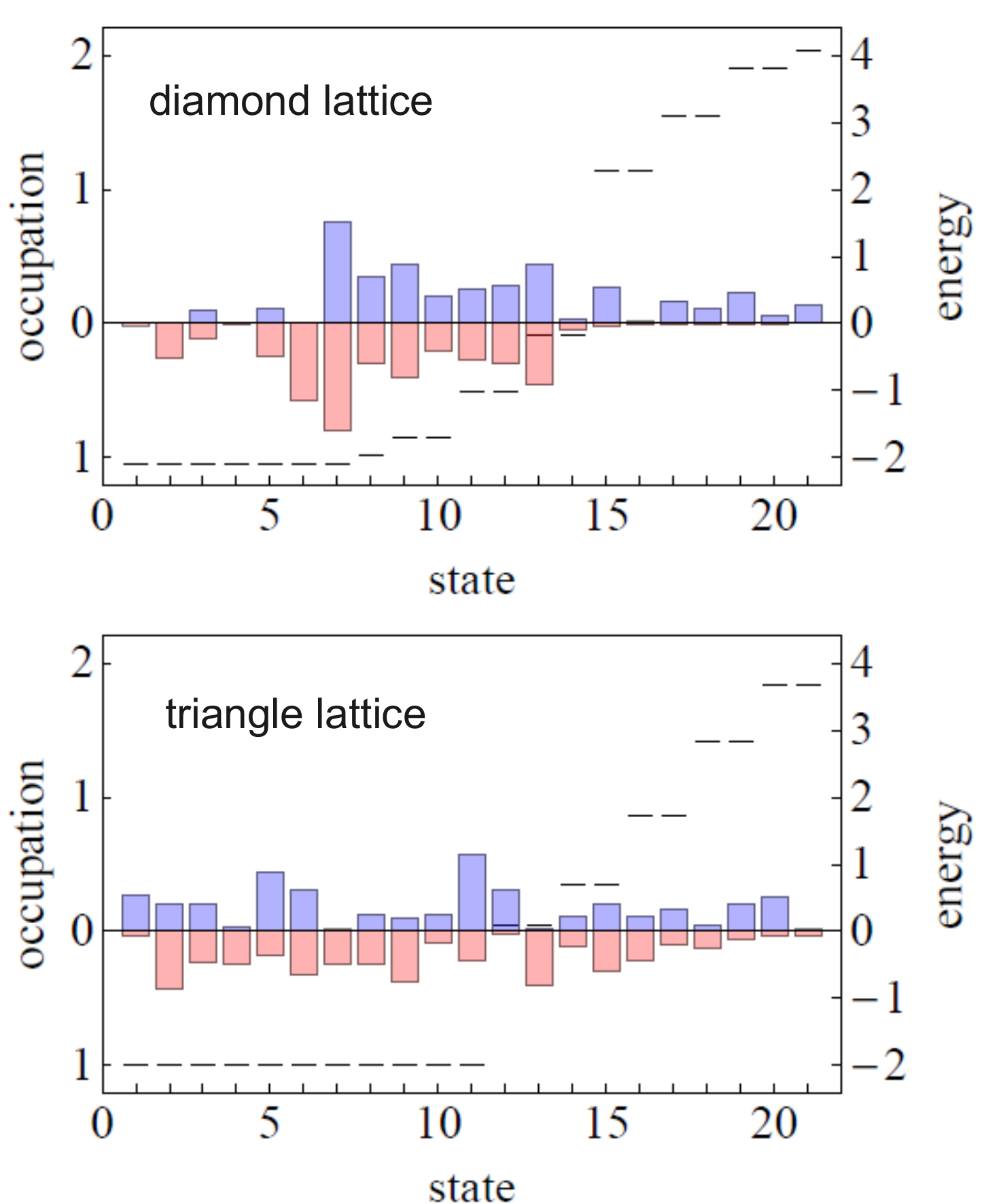}
\caption{(Color online) Occupation of single particle states in the
  initial many-body state (in the presence of the harmonic
  confinement) for diamond and triangle lattices. The single particle
  energy levels (without the confinement) are shown as short lines and
  the corresponding occupancies as bars, above the horizontal line
  (blue) for bosons and below (red) for fermions. Here the flat band
  has the lowest energy.}
\label{occdist}
\end{figure}

\begin{figure}[h!]
\includegraphics[width=0.7\columnwidth]{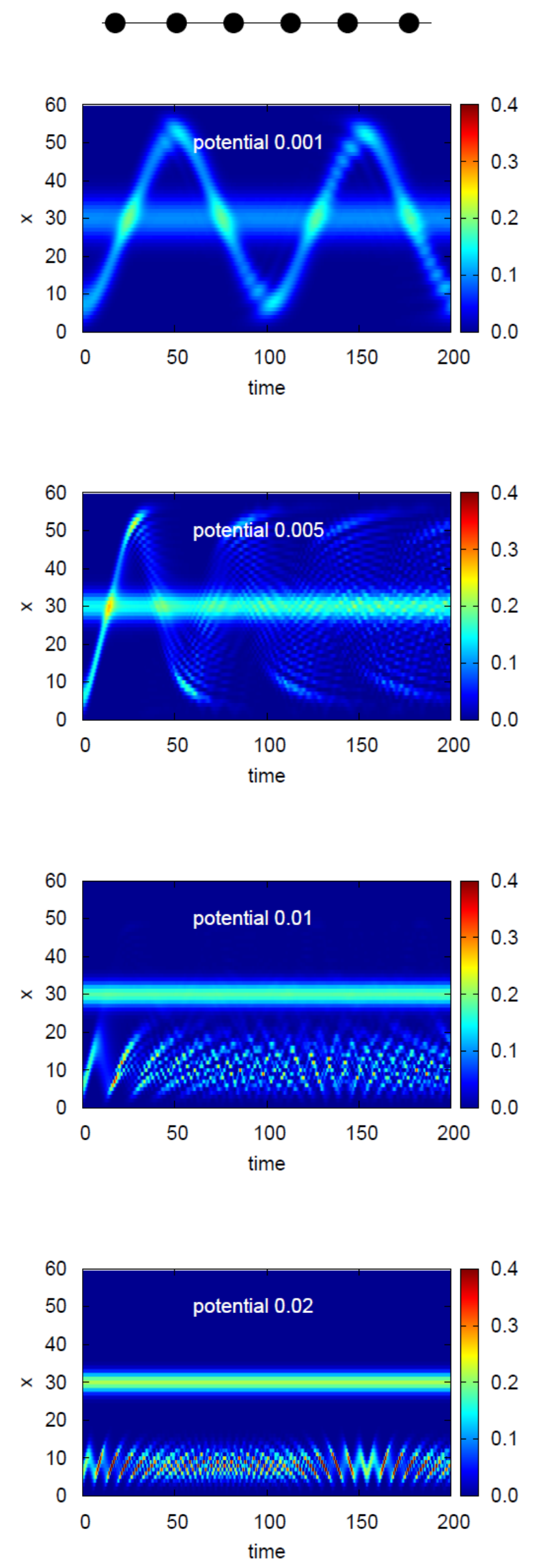}
\caption{(Color online) Attempts to make a Newton's cradle with two harmonic
  potentials easily end up creating two separate clouds due to Bloch
  oscillations. The figures represent a simple linear chain density of
  two bosons originally in the ground state of a double harmonic well
  potential. They are set in motion by leaving only one minimum, with
  potential $V_h(i)=\alpha i^2$, where $i$ is the site index and the
  strength $\alpha$ as indicated in the figures. A shallow potential
  (upper left figure) sustains Newton's cradle oscillations for over
  10 cycles. Time is measured in units of inverse energy ($\hbar=1$).
}
\label{linearcradle}
\end{figure}

\begin{figure}[h!]
\includegraphics[width=.8\columnwidth]{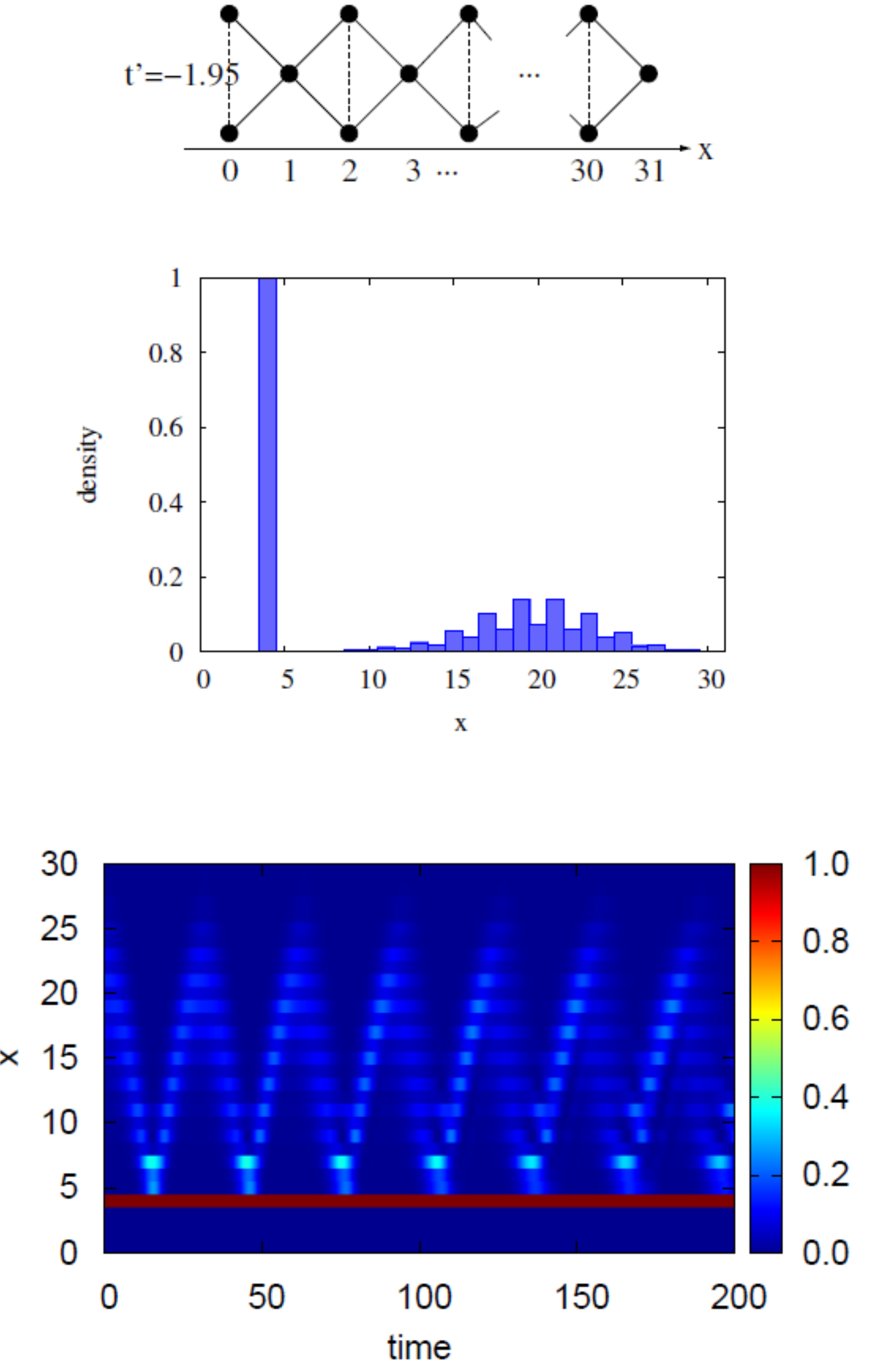}
\caption{(Color online) The upmost figure shows the diamond chain and the definition 
of the linear coordinate $x$. The next figure shows the chosen initial density,
with one atom on the left on the flat band, and the other one being less localized.  
After leaving only a shallow harmonic potential with minimum at $x=4$ the cloud on 
the right is driven left. The time evolution of the density is shown in the lowest figure,
which shows how the occupied, low-energy flat band acts in this case as a reflective wall.}
\label{diamondcradle}
\end{figure}

In the cases where the flat band is the lowest band the initial boson
and fermion distributions are slightly different due to the different
statistics. In order to get more insight into the initial many-body
state we determined the single particle state occupancies for bosons and
fermions using Eq.~(\ref{statedist}). The single particle states are the
tight-binding eigenstates without the confining potential.
The results are shown in Fig.~\ref{occdist}.
The figure shows that the many-body state of the Bose system has
nearly as large occupancy of the flat-band states as the Fermi
system. In fermion systems those particles are immobile, but in boson
systems they interact with the particles in the normal band, and
become mobile.

\subsection{Newton's cradle and Bloch oscillations}

As we saw, depending on the occupation of states flat-band systems can
have both inert and mobile electrons. In principle this could resemble
a quantum Newton's cradle, demonstrated in a trapped Bose
gas by Kinoshita {\it et al.} \cite{kinoshita-2006}.  Ideally there
is no dephasing, and perfect momentum exchange among indistinguishable
atoms can leave some atoms immobile, just like those on a flat band. A
practical limitation to the realization of a Newton's cradle with atoms
in a discrete lattice is posed by Bloch oscillations \cite{bloch-29}:
If one drives atoms with a too steep potential slope, they accelerate
only to the point where their group velocity is so high that a Bragg
reflection reverses their motion. 

Let us look first at a basic linear chain. In Fig.~\ref{linearcradle}
we have two atoms originally in the ground state of two harmonic
potentials, with minima at sites 6 and 30. The plotted density scale
is chosen to be from 0 to 0.4, which covers the cases studied here.
Subsequently we remove the
minimum at 6 and follow the time evolution of the density. The
amplitude of the trapping potentials determines how localized the
atoms are originally. The details of the initial states are not
important, but after removing the minimum at 6 the atom cloud around
site 6 sits on a potential slope, the steepness of which is determined
by the amplitude. With a shallow potential we get a nice Newton's cradle
where oscillations are only slightly distorted and damped.
The damping is caused by many-body
effects. If the potential slope is made too steep, however, Bloch
oscillations set in before atom clouds have a chance to collide, and
one can even have a rapidly oscillating cloud with no interference
with the other cloud. In these examples Bloch oscillations are
symmetric, while in general they can be asymmetric \cite{dahan-1996}.

Next we try something similar in a flat-band system. A suitably mutable
flat-band system is the diamond chain with the vertical hopping
parameter that shifts the flat-band level (see fourth panel in
Fig.~\ref{fig1}). Here the flat-band energy is chosen to be almost
degenerate with the low edge of the lowest normal band. We put the
potential minima at $x=4$ and $x=20$, where $x$ is the coordinate 
along the diamond chain.
In order to see the effect of the flat band, we want about 50
percent occupation on it.  This is achieved by pinning one atom with a
very strong potential to $x=4$, which in our chain has two sites (at
the tips of the diamond). We want the other atom to remain mobile, so
we use a shallow potential with amplitude 0.001 to put it around
$x=20$, but not on the flat band. Next we wish to set the latter atom
in motion toward the first atom. The driving potential has to be very
shallow to avoid Bloch oscillations. We use again a
harmonic potential with amplitude 0.001, which for the linear chain
produced a nice Newton's cradle, with the minimum at $x=4$.

The result is presented in Fig~\ref{diamondcradle} - the plotted
density ranges now from 0 to 1 because there is exactly one atom on
the flat band.  Instead of a Newton's cradle we get a reflection from
the flat-band atoms at $x=4$. These atoms are now held in place, not
by a potential, but by the flatness of the band. Such reflection is
peculiar to flat-band systems, a mere confinement as in the linear
chain case gives rise to momentum exchange in a Newton's cradle
fashion. Also the unoccupied flat-band states at low energy affect the
motion of atoms: the horizontal stripes in the oscillating part are
density maxima at the node points of the diamond lattice (odd $x$
values).  This is simply a consequence of the fact that while the atoms
occupying a flat band have now no way of leaving the band and moving,
there is also no way the moving atoms can occupy a flat band. As a
by-product, one could create a confining box in 1D by pinning two
atoms to two well-separated even-$x$ sites (which puts them on a flat
band). Other bosons in between would then be trapped by the two mirrors.

\subsection{Transport}
\begin{figure}[h!]
\includegraphics[width=0.75\columnwidth]{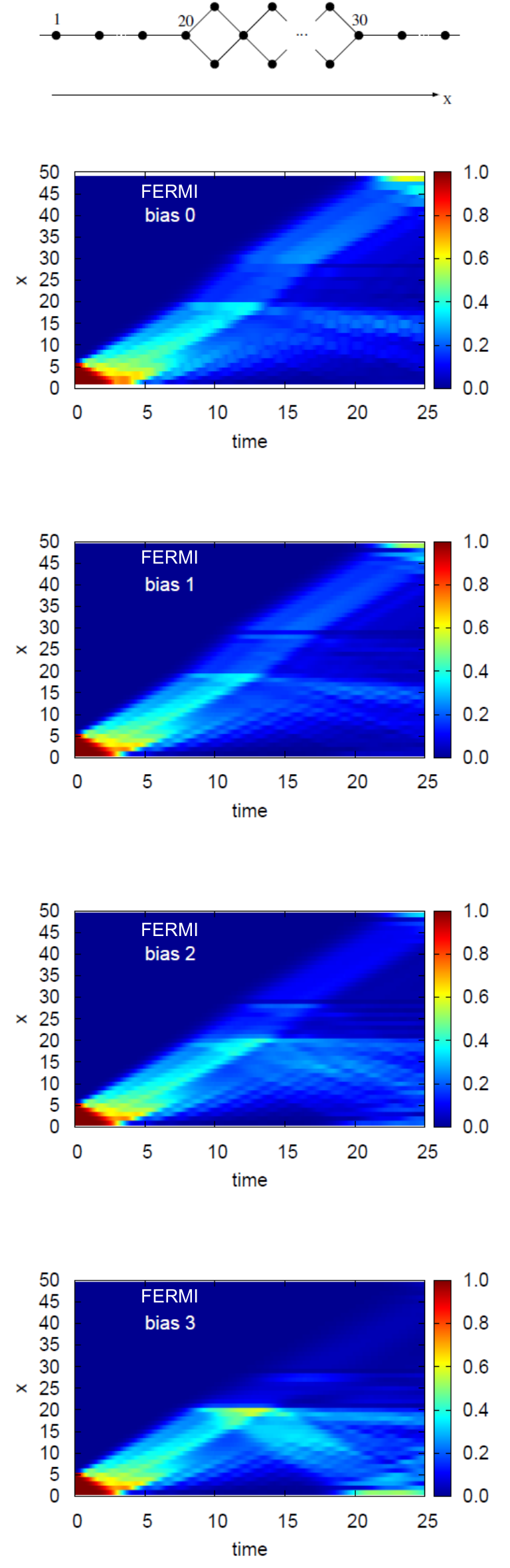}
\caption{(Color online) A diamond chain between two finite leads. Five
  fermions are initially on the leftmost sites and the 
  potentials of the leads have been shifted by the bias value. The diamond
  chain is in the range $20<x<30$. The calculation was done using TEBD with
  100x100 matrix product states.}
\label{leads}
\end{figure}

The transport properties of a delta chain between semi-infinite leads
have been investigated in Ref.~\cite{schulze-bercioux-urban-2012} by
Schulze {\it et al.} in the case of adiabatic pumping, {\it i.e.}
without external bias.  In Fig.~\ref{leads} we show an example of the
time evolution in a flat-band system, now a diamond chain, between
finite leads. Here the diamond lattice has normal bands symmetrically
above and below the flat band (see Fig.~\ref{fig1}). The potential
difference between the left lead and the diamond lattice is given by
the bias (e.g. for electrons a voltage bias). For a small bias,
fermions can occupy the low-energy normal band of the diamond chain and
some reflect from the end of the the diamond chain. For a large bias
the potential step dominates and fermions reflect. In this setup
there is little difference between bosons and fermions.

\begin{figure}[h!]
\includegraphics[width=0.75\columnwidth]{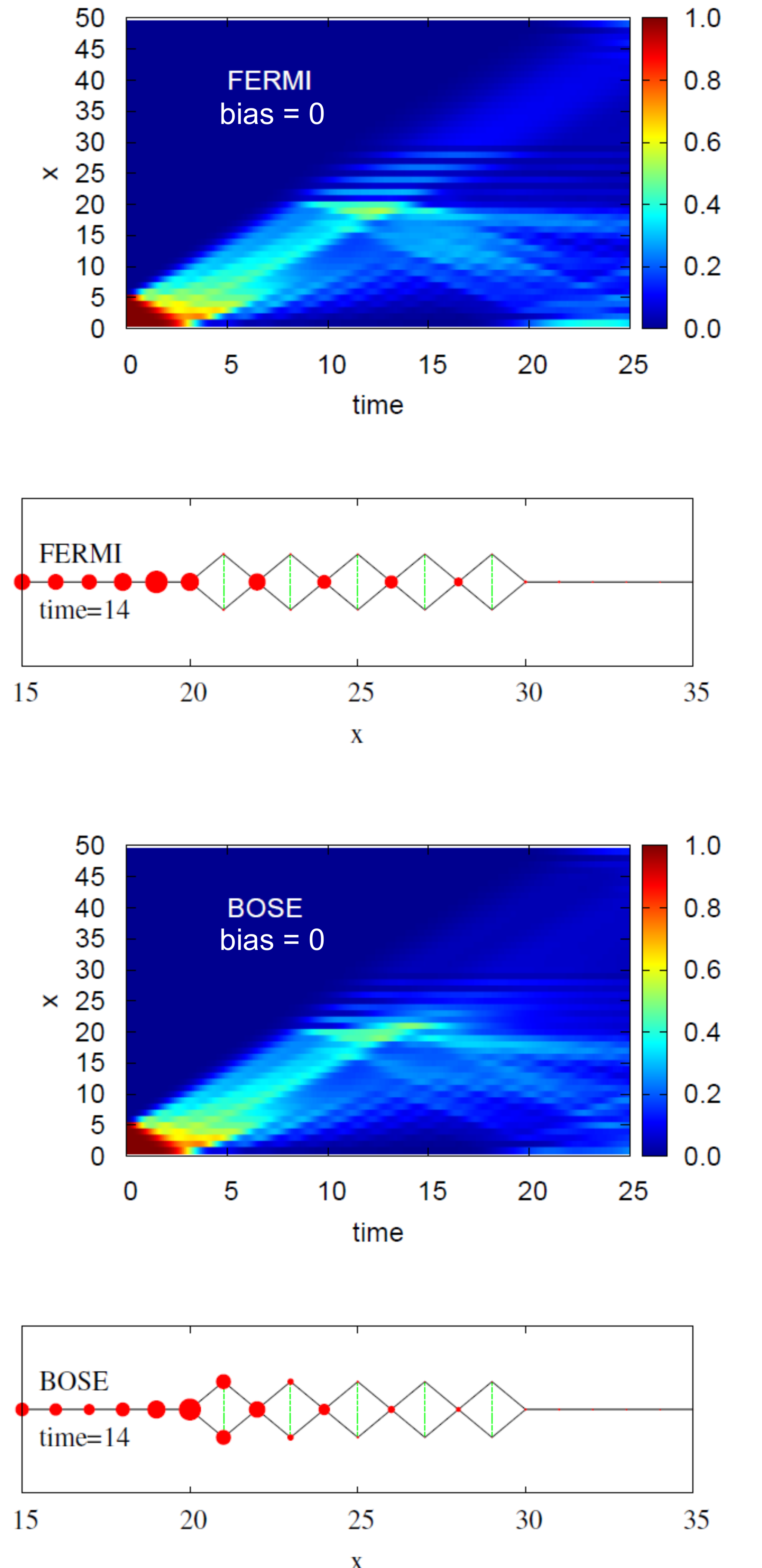}
\caption{(Color online) A diamond chain with transverse hopping between two finite leads.
  Apart from the modified band structure, the setup is the same as in Fig.~\ref{leads}.  
  At time=0 the five bosons or fermions are at lead sites 1 to 5, and are set in motion
  by diffusion. The upper figures show the time evolution of the density for fermions 
  and the lower figures for bosons. Snapshots at time=14 show how sites are 
  occupied by fermions and bosons; the circle size is proportional to the density at that site.   
  Differences arise from the boson {\it vs.} fermion statistics. 
}
\label{leads-bose-fermi}
\end{figure}

The flat band can be shifted up or down in energy with respect to the
normal bands by adding a transverse hopping $t'$, as visualized in
Fig.~\ref{fig1}. Using the value $t'= -2$ the flat band is pushed down
to become the lowest band, degenerate with the lower edge of the lower
dispersive band.  Fig.~\ref{leads-bose-fermi} shows again the
time-dependent density. There is no bias, particle motion is now
driven by diffusion.

To interpret Fig.~\ref{leads-bose-fermi} we note that the amplitude of
the flat-band, single-particle state, which cannot be occupied by moving
particles, is highest at the tips of the diamond (at $x=21,23,...29$).
This throttles the motion of the particles, as they must use
the normal band degenerate with the flat band. Fermions are affected more
because of the Pauli principle (or the fermion phase factor,
if one thinks of the current). Fermion density at the tips of the
diamond remains always low, and they pile up in the nodal sites.  This
causes the horizontal stripes in the fermion density in
Fig.~\ref{leads-bose-fermi}, and is seen clearly in the snapshot
at time 14 of the density at each site, which is shown in the same figure.

Up to approximately time 10 bosons diffuse to the diamond lattice
almost exactly like fermions: they pile up in the nodal site at
$x=22$, and tip sites remain almost empty.  As mentioned above, the
low-energy normal band is degenerate with the flat band.  Boson
occupation of these low-energy dispersive states is not limited, and
bosons begin to fill the tip sites at $x=21$.  This is clearly seen in
the snaphot at time 14 in the lowest panel of
Fig.~\ref{leads-bose-fermi}. Due to the deflection to the tip sites
the diffusion of bosons through the diamond lattice is slowed down.

\section{Conclusions}

We have addressed the differences in particle dynamics between bosons
and fermions in quasi-one-dimensional lattices with a flat band. The
particles were assumed to be spinless and interacting with an
infinitely strong contact interaction.  In this case the fermions are
equivalent to noninteracting particles due to the Pauli exclusion
principle.  Consequently, the fermions occupying the flat-band states
do not contribute to the persistent current in a quantum ring, and
they are localized in the lattice.

In the case of bosons the particles are truly interacting making the system
more interesting. The persistent current shows a particle-hole symmetry
and the occupation of the flat band can also change the current.
The bosons do not stay localized in the lattice
even if initially the partition of the many-body wave function to the single particle
states is essentially the same as in the case of fermions.

These results are very general, they do not seem to depend on the
detailed structure of the lattice or on the position of the flat band
with respect to the normal bands with dispersion. This gains in
importance only when the flat-band lattice is brought in contact with an
external lattice.

{\bf Acknowledgments}: Ideas for this work were initiated by a
 workshop of the European Science Foundation research networking 
program POLATOM.
We thank the Academy of Finland for financial support.

\vskip10pt

\bibliographystyle{unsrt}
\bibliography{flatpaper2.bib}

\begin{thebibliography}{10}

\bibitem{stewart1984}
G~R Stewart.
\newblock {\em Rev. Mod. Phys.}, {56}:755, 1984.

\bibitem{syozi1951}
I~Syo\^zi.
\newblock {\em Prog. Theor. Phys.}, 6:306, 1951.

\bibitem{deng2003}
S~Deng, A~Simon, and J~K\"ohler.
\newblock {\em J. Solid State Chem.}, 176:412, 2003.

\bibitem{miyahara2005}
S~Miyahara, K~Kubo, H~Ono, Y~Shimomura, and N~Furukawa.
\newblock {\em J. Phys. Soc. Japan}, 74:1918, 2005.

\bibitem{castroneto2009}
A~H~Castro Neto, F~Guinea, N~M~R Peres, K.S. Novoselov, and A.K. Geim.
\newblock {\em Rev. Mod. Phys.}, 81:109, 2009.

\bibitem{lee2001}
H~Lee, J~A Johnson, M~Y He, J~S Speck, and P~M Petroff.
\newblock {\em Appl. Phys. Lett.}, {78}:105, 2001.

\bibitem{koskinen2003}
M~Koskinen, S~M Reimann, and M~Manninen.
\newblock {\em Phys. Rev. Lett.}, {90}:066802, 2003.

\bibitem{reimann2002}
S~M Reimann and M~Manninen.
\newblock {\em Rev. Mod. Phys}, {74}:1283, 2002.

\bibitem{bloch2005}
I~Bloch.
\newblock {\em Nature Physics}, 1:23, 2005.

\bibitem{leggett2006}
A~J Legget.
\newblock {\em Quantum Liquids: Bose condensation and Cooper pairing in
  condensed matter physics}.
\newblock Oxford University Press, 2006.

\bibitem{pethick2008}
C~J Pethick and H~Smith.
\newblock {\em Bose-Einstein Condensation in Dilute Gases}.
\newblock Cambridge University Press, 2 edition, 2008.

\bibitem{lopezaguayo2010}
S~L\'opez-Aguayo, Y~V Kartashov, V~A Vysloukh, and L~Torner.
\newblock {\em Phys. Rev. Lett.}, 105:013902, 2010.

\bibitem{bloch2008}
I~Bloch.
\newblock {\em Science}, {319}:1202, 2008.

\bibitem{bloch2008b}
I~Bloch, J~Dalibard, and W~Zwerger.
\newblock {\em Rev. Mod. Phys.}, 80:885--964, 2008.

\bibitem{struck-etal-2011}
J.~Struck, C.~\"Olschl\"ager, R.~Le Targat, P.~Soltan-Panahi, A.~Eckardt,
  M.~Lewenstein, P.~Windpassinger, and K.~Sengstock1.
\newblock Quantum simulation of frustrated classical magnetism in triangular
  optical lattices.
\newblock {\em Science}, 333(6045):996--999, 2011.

\bibitem{jo-2012}
Gyu-Boong Jo, Jennie Guzman, Claire~K. Thomas, Pavan Hosur, Ashvin Vishwanath,
  and Dan~M. Stamper-Kurn.
\newblock Ultracold atoms in a tunable optical kagome lattice.
\newblock {\em Phys. Rev. Lett.}, 108:045305, Jan 2012.

\bibitem{apaja2010}
V~Apaja, M~Hyrk\"as, and M~Manninen.
\newblock {\em Phys. Rev. A}, {82}:041402, 2010.

\bibitem{kinoshita-2006}
Toshiya Kinoshita, Trevor Wenger, and David~S. Weiss.
\newblock A quantum newton's cradle.
\newblock {\em Nature}, 440(7086):900--903, Apr 2006.

\bibitem{saarikoski2010}
H~Saarikoski, S~M Reimann, A~Harju, and M~Manninen.
\newblock {\em Rev. Mod. Phys.}, {82}:2785, 2010.

\bibitem{paredes-2004}
Belen Paredes, Artur Widera, Valentin Murg, Olaf Mandel, Simon Folling, Ignacio
  Cirac, Gora~V. Shlyapnikov, Theodor~W. Hansch, and Immanuel Bloch.
\newblock Tonks-girardeau gas of ultracold atoms in an optical lattice.
\newblock {\em Nature}, 429(6989):277--281, May 2004.

\bibitem{vidal-2000}
Julien Vidal, Benoit Dou\ifmmode~\mbox{\c{c}}\else \c{c}\fi{}ot, R\'emy
  Mosseri, and Patrick Butaud.
\newblock Interaction induced delocalization for two particles in a periodic
  potential.
\newblock {\em Phys. Rev. Lett.}, 85:3906--3909, Oct 2000.

\bibitem{doucot-2002}
Benoit Dou\ifmmode~\mbox{\c{c}}\else \c{c}\fi{}ot and Julien Vidal.
\newblock Pairing of cooper pairs in a fully frustrated josephson-junction
  chain.
\newblock {\em Phys. Rev. Lett.}, 88:227005, May 2002.

\bibitem{vidal2004}
G~Vidal.
\newblock {\em Phys. Rev. Lett.}, {93}:040502, 2004.

\bibitem{viefers2004}
S~Viefers, P~Koskinen, P~Singha Deo, and M~Manninen.
\newblock {\em Physica E}, { 21}:1, 2004.

\bibitem{reppy1964}
J~D Reppy and D~Depatie.
\newblock {\em Phys. Rev. Lett.}, {12}:187, 1964.

\bibitem{grobman1996}
W~Grobman and M~Luban.
\newblock {\em Phys. Rev.}, {147}:166, 1966.

\bibitem{lyandageller2000}
Y~Lyanda-Geller and P~M Goldbart.
\newblock {\em Phys. Rev. A}, {61}:043609, 2000.

\bibitem{isoshima2000}
T~Isoshima, M~Nakahara, T~Ohmi, and K~Machida.
\newblock {\em Phys. Rev. A}, {61}:063610, 2000.

\bibitem{wang2007}
T~Wang, J~Javanainen, and S~F Yelin.
\newblock {\em Phys: Rev. A}, {76}:011601, 2007.

\bibitem{ryu2007}
C~Ryu, M~F Andersen, P~Clad\'e, V~Natarajan, K~Helmerson, and W~D Phillips.
\newblock {\em Phys Rev Lett}, 99:260401, 2007.

\bibitem{bargi2010}
S~Bargi, F~Malet, G~M Kavoulakis, and S~M Reimann.
\newblock {\em Phys. Rev. A}, {82}:043631, 2010.

\bibitem{lembessis2010}
V~E Lembessis and M~Babiker.
\newblock {\em Phys. Rev. A}, {82}:051402, 2010.

\bibitem{hopkins2004}
A~Hopkins, B~Lev, and H~Mabuchi.
\newblock {\em Phys. Rev. A}, {70}:053616, 2004.

\bibitem{morizot2006}
O~Morizot, Y~Colombe, V~Lorent, H~Perrin, and B~M Garraway.
\newblock {\em Phys. Rev. A}, {74}:023617, 2006.

\bibitem{griffin2008}
P~F Griffin, E~Riis, and A~S Arnold.
\newblock {\em Phys. Rev. A}, {77}:051402, 2008.

\bibitem{baker2009}
P~M Baker, J~A Stickney, M~B Squires, J~A Scoville, E~J Carlson, W~R Buchwald,
  and S~M Miller.
\newblock {\em Phys. Rev. A}, {80}:063615, 2009.

\bibitem{mueller2004}
E~J Mueller.
\newblock {\em Phys. Rev. A}, {70}:041603, 2004.

\bibitem{amico2005}
L~Amico, A~Osterloh, and F~Cataliotti.
\newblock {\em Phys. Rev. Lett.}, {95}:063201, 2005.

\bibitem{fetter2009}
A~L Fetter.
\newblock {\em Rev. Mod. Phys.}, {81}:647, 2009.

\bibitem{lieb1963}
E~H Lieb and W~Liniger.
\newblock {\em Phys Rev}, 130:1605, 1963.

\bibitem{lieb1963b}
E~H Lieb.
\newblock {\em Phys Rev}, 130:1616, 1963.

\bibitem{yang1967}
C~N Yang.
\newblock {\em Phys Rev Lett}, 19:1312, 1967.

\bibitem{lieb1968}
E~H Lieb and F~Y Wu.
\newblock {\em Phys. Rev. Lett.}, {20}:1445, 1968.

\bibitem{manninen-reimann-viefers-2012}
M.~Manninen, S.~M. Reimann, and S.~Viefers.
\newblock Quantum rings for beginners ii: Bosons versus fermions.
\newblock {\em Physica E}, 46:119, 2012.

\bibitem{bloch-29}
F.~Bloch.
\newblock {\em Z. Phys.}, 52:555, 1929.

\bibitem{dahan-1996}
Maxime Ben~Dahan, Ekkehard Peik, Jakob Reichel, Yvan Castin, and Christophe
  Salomon.
\newblock Bloch oscillations of atoms in an optical potential.
\newblock {\em Phys. Rev. Lett.}, 76:4508--4511, Jun 1996.

\bibitem{schulze-bercioux-urban-2012}
M.~Schulze, D.~Bercioux, and D.~F. Urban.
\newblock Adiabatic pumping in the quasi-one-dimensional triangle lattice.
\newblock {\em arXiv:1208.6113v1}, 2012.

\end{thebibliography}

\end{document}